# A Novel Chaotic Image Encryption using Generalized Threshold Function

Sodeif Ahadpour
Department of Sciences,
University of Mohaghegh
Ardabili, Ardabil, Iran.
ahadpour@uma.ac.ir

Yaser Sadra
Department of Sciences,
University of Mohaghegh
Ardabili, Ardabil, Iran.
sadra@staff.um.ac.ir

Zahra ArastehFard
Department of Sciences,
University of Mohaghegh
Ardabili, Ardabil, Iran.
z.arastehfard@msn.com

## ABSTRACT
*In this paper, after reviewing the main points of image encryption and threshold function, we introduce the methods of chaotic image encryption based on pseudorandom bit padding that the bits be generated by the novel generalized threshold function (segmentation and self-similarity) methods. These methods decrease periodic effect of the ergodic dynamical systems in randomness of the chaotic image encryption. The essential idea of this paper is that given threshold functions of the ergodic dynamical systems. To evaluate the security of the cipher image of this scheme, the key space analysis, the correlation of two adjacent pixels and differential attack were performed. This scheme tries to improve the problem of failure of encryption such as small key space and level of security.*

## 1. INTRODUCTION
In recent years, the cryptographic schemes schemes have suggested some new and efficient ways to develop secure image encryption [4]. These schemes have typical structure which performed the permutation and the diffusion stages, alternatively. However, most of algorithms be faced with some problems such as the lack of robustness and security. The random number generators are intransitive in cryptography for generation of cryptographic keys, allegorically, secret keys utilized in symmetric cryptosystems [1,2] and large numbers is intransitive in asymmetric cryptosystems [3,5], because of unpredictable, should better be generated randomly. In addition, random number generators in many cryptographic protocols, such as to create challenges, blinding values, Monte Carlo methods are used [6,7,8]. Also, the random number generators are used more in the diffusion functions of the image encryption for diffused pixels of plain image.

Random number generators can be classified into three classes which are pseudorandom number generators (PRNGs), true random number generators (TRNGs) and hybrid random number generators (HRNGs). PRNGs use deterministic processes to generate a series of outputs from an initial seed state [9,10,11]. TRNGs use of non-deterministic source (i.e., the entropy source), along with some processing function (i.e., the entropy distillation process) to generate the random bit sequence [1]. These sources consist of physical phenomena such as atmospheric noise, thermal noise, radioactive decay and even coin-tossing [12]. Many PRNGs using chaotic maps have been established. Most of them have very complex structures. In this paper, we propose a chaotic image encryption based on pseudorandom bit padding that the bits be generated by the novel generalized threshold function (segmentation and self-similarity) methods. The random bit sequences produced by this generator are evaluated using the 15 statistical tests recommended by U.S. NIST [1]. Experimental results show that this PRNG possess good uniformity and randomness properties.

This paper is arranged as follows. In section 2, the properties of the threshold functions of ergodic dynamical systems are discussed. In section 3, we introduce the proposed random number generators and then discuss the uniformity and randomness of the bit sequences generated by the Proposed PRNG. In section 4, we propose chaotic encryption scheme based on pseudorandom bit padding and finally, in Section 5, we conclude the paper.

## 2. THRESHOLD FUNCTIONS OF ERGODIC DYNAMICAL SYSTEMS
*One of the easiest and most widely algorithms for generating chaotic encryption scheme, used of the threshold function [13,14]. Threshold function is a function that takes the value 0 if a specified function of the arguments exceeds a given threshold and 1 otherwise. To consider a one-dimensional chaotic system which is defined as follows:*

$$x_n = f(x_{n-1}) \qquad n = 0,1,2,...$$

*that $f : I \to I$ ( $I = [0,1]$) is a nonlinear map. Threshold function of ergodic dynamical system is approximately the middle of the minimum and maximum of the $\{x_n\}_{n=0}^{\infty}$ values. In the other hand, threshold function of ergodic dynamical system can be defined as follows:*

$$c = \int_0^1 f(x) f^*(x) dx \qquad (1)$$

*where the $f^*$ is density function of the nonlinear map $f$ [15]. One of the methods of generating is as follows, to consider above one-dimensional chaotic system, the* pseudorandom bit sequence $\{z_n\}_{n=0}^{\infty}$ can be defined as follows (see Fig.1(a)):

$$z_n = \begin{cases} 0 & x_n < c \\ 1 & x_n \geq c \end{cases} \quad (2)$$

that $c$ is threshold function of the $\{x_n\}_{n=0}^{\infty}$ values. In other words, threshold function divides the interval of I=[0,1] into 2 boxes. Widths of boxes are ($c$) and ($1-c$), respectively.

## 3. THE PROPOSED PRNG AND RANDOMNESS ANALYSIS

In this section, we introduce two proposed pseudorandom number generators based on the generalized threshold function and then their randomness is discussed.

### 3.1 The proposed PRNG based on segmentation method

To consider a one-dimensional chaotic system which is defined as follows:

$$x_n = f(x_{n-1}) \quad n = 0,1,2,...$$

that $f : I \rightarrow I$ ($I = [0,1]$) is a nonlinear map. In this method, we divided the interval of I=[0,1] into $2^{k-1}$ blocks and then divided each block into 2 boxes by threshold function (i.e., $c$). Consequently, we have $2^k$ boxes in the interval of I=[0,1]. Width of boxes are ($c \times \frac{1}{2^{k-1}}$) and (($1-c) \times \frac{1}{2^{k-1}}$), respectively. In this case, the pseudorandom bit sequence $\{z_n\}_{n=0}^{\infty}$ is defined as follows (see Fig.1(b)):

$$z_n = \begin{cases} 0 & [x_n \times 2^{k-1}] \leq c \\ 1 & other\ wise \end{cases} \quad (3)$$

where symbol of [] is the fractional part. With this way, randomness of the PRNG is increased, because, this method decreased periodic effect of the chaotic maps in randomness of the PRNG. With regard to the above contents, we can be redefined this method in two-dimension. For this purpose, to consider a two-dimensional chaotic system which is defined as follows:

$$x_n = f(x_{n-1}) \quad n = 0,1,2,...$$
$$y_n = g(y_{n-1}) \quad n = 0,1,2,...$$

that $f$ and $g : I \rightarrow I$ ($I = [0,1]$) are nonlinear maps. Threshold functions of the $\{x_n\}_{n=0}^{\infty}$ values and the $\{y_n\}_{n=0}^{\infty}$ values are $c$ and $c'$, respectively. Therefore, the pseudorandom bit sequence $\{z_n\}_{n=0}^{\infty}$ is defined as follows (see Fig.1(c)):

$$z_n = \begin{cases} 0 & (A\ or\ B) \\ 1 & other\ wise \end{cases} \quad (4)$$

Where,

$A \equiv ([x_n \times 2^{k-1}] \leq c$ and $[y_n \times 2^{k-1}] \leq c')$
$B \equiv ([x_n \times 2^{k-1}] > c$ and $[y_n \times 2^{k-1}] > c')$

and symbol of [] is the fractional part. As an example, we consider the chaotic logistic map

$$x_n = r_x x_{n-1}(1 - x_{n-1})$$

for the functions of $f$ and $g$ ($x_n, y_n \in (0,1)$ and $r_x, r_y \in (3.99996, 4])$ [16]. In this case, threshold functions are $\frac{1}{2}$ (i.e. $c = c' = \frac{1}{2}$) [17].

Thus, pseudorandom bit sequence $\{z_n\}_{n=0}^{\infty}$ can be defined as follows Eq. 4, Where,

$A \equiv ([x_n \times 2^{k-1}] \leq \frac{1}{2}$ and $[y_n \times 2^{k-1}] \leq \frac{1}{2})$
$B \equiv ([x_n \times 2^{k-1}] > \frac{1}{2}$ and $[y_n \times 2^{k-1}] > \frac{1}{2})$

and symbol of [] is the fractional part.

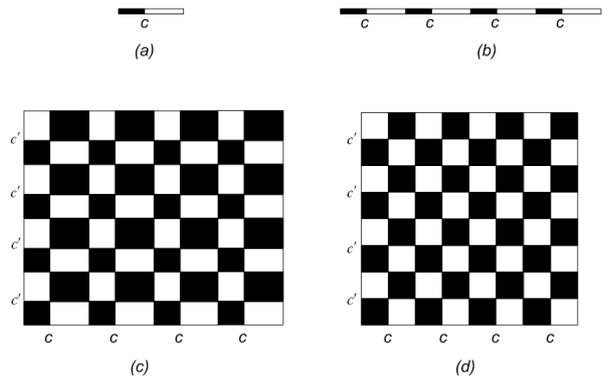

Fig 1: a) threshold function of a block, (b) the generalized threshold function of $2^2$,(k=3) blocks in one-dimension, (c) the generalized threshold function of $2^2$,(k=3) blocks in two-dimension, (d) the generalized threshold function of $2^2$,(k=3) blocks in two-dimension for logistic map (1), the boxes of 0 and 1 be showed black and white colors, respectively.

### 3.2 The proposed PRNG based on self-similarity method

Self-similarity is the property that a substructure to be analogous or identical with an overall structure [18]. In the other words, a self-similar object behaves the same when viewed at different degrees of magnification, or different scales on a dimension (space or time) [18]. As we have mentioned, threshold function is a function that takes the value 0 if a specified function of the arguments exceeds a given threshold and 1 otherwise. In this method, first, to consider a one-dimensional chaotic system which is defined as follows:

$$x_n = f(x_{n-1}) \quad n = 0,1,2,...$$

*2*

that $f : I \to I$ ($I = [0,1]$) is a nonlinear map. Next, we consider this interval as a block and divide this block into 2 boxes by threshold function (i.e. $[0,c_1],(c_1,1]$) (see Fig. 2(a)). Then, we consider again each of the boxes as a block and divide each of these blocks into 2 boxes by threshold function (i.e. $[0,c_{21}],(c_{21},c_1],(c_1,c_1+c_{22}],(c_1+c_{22},1]$) (see Fig. 2(b)). We iterate this divisions k times. In each iteration, threshold function showed with $c_{ij}$ that index of $i$ indicant $i$ th iteration and index of $j$ indicant $j$ th block. Consequently, the interval of I=[0,1] divided into $2^k$ boxes by k times iteration of threshold function and the interval of I=[0,1] converts as follows:

$$I = [0,c_{k1}] \cup (c_{k1},c_{(k-1)1}] \cup (c_{(k-1)1}, c_{(k-1)1}+c_{k2}]$$
$$\cup ... \cup (c_1+...+c_{k2^{k-1}}, 1].$$

In the other hand, this interval using threshold function has property of self-similarity. Self-similarity in an interval is the property that a subinterval to be analogous or identical with an overall interval (see Fig. 2(b)). In here, degree of Self-similarity is dependent to the number of above iterations, i.e. if the interval divided into $2^k$ boxes, degree of Self-similarity is k. In this case, the pseudorandom bit sequence $\{z_n\}_{n=0}^{\infty}$ is defined as follows (see Fig.2(b)):

$$z_n = \begin{cases} 0 & A \\ 1 & other\ wise \end{cases} \quad (5)$$

Where,

$A \equiv x_n \leq c_{k1}$ or $c_{(k-1)1} < x_n \leq c_{(k-1)1}+c_{k2}$ or
... or $c_1+...+c_{(k-1)2^{k-2}} < x_n \leq c_1+...+c_{k2^{k-1}}$.

With this method, randomness of this PRNG is higher than the previous PRNG, because, higher than the previous method, this method decreased periodic effect of the chaotic maps in randomness of the PRNG. With regard to the above contents, we can be redefined this method in two-dimension. For this purpose, to consider a two-dimensional chaotic system which is defined as follows:

$$x_n = f(x_{n-1}) \quad n = 0,1,2,...$$
$$y_n = g(y_{n-1}) \quad n = 0,1,2,...$$

that $f$ and $g : I \to I$ ($I = [0,1]$) are nonlinear maps. Threshold functions of the $\{x_n\}_{n=0}^{\infty}$ values and the $\{y_n\}_{n=0}^{\infty}$ values are c and $c'$, respectively. Therefore, the pseudorandom bit sequence $\{z_n\}_{n=0}^{\infty}$ is defined as follows (see Fig.2(c)):

$$z_n = \begin{cases} 0 & (A\ or\ B) \\ 1 & other\ wise \end{cases} \quad (6)$$

Where,

$A \equiv [(x_n \leq c_{k1}$ or $c_{(k-1)1} < x_n \leq c_{(k-1)1}+c_{k2}$ or
... or $c_1+...+c_{(k-1)2^{k-2}} < x_n \leq c_1+...+c_{k2^{k-1}})$ and
$(y_n \leq c'_{k1}$ or $c'_{(k-1)1} < y_n \leq c'_{(k-1)1}+c'_{k2}$ or
... or $c'_1+...+c'_{(k-1)2^{k-2}} < y_n \leq c'_1+...+c'_{k2^{k-1}})]$

$B \equiv [(c_{k1} < x_n \leq c_{(k-1)1}$ or $c_{(k-1)1}+c_{k2} < x_n \leq c_{(k-1)2}$ or ... or $c_1+...+c_{k2^{k-1}} < x_n \leq 1)$ and
$(c'_{k1} < y_n \leq c'_{(k-1)1}$ or $c'_{(k-1)1}+c'_{k2} < y_n \leq c'_{(k-1)2}$ or ... or $c'_1+...+c'_{k2^{k-1}} < y_n \leq 1)]$.

As an example, we consider one of the one-parameter families of chaotic maps [19]

$$x_n = \frac{\alpha_x^2 (2x_{n-1}-1)^2}{4x_{n-1}(1-x_{n-1}) + \alpha_x^2 (2x_{n-1}-1)^2}$$

for the functions of $f(x_n \in (0,1)$ and $\alpha = 0.75)$ and $g(y_n \in (0,1)$ and $\alpha = 1.5)$ [19]. In this case, according to the Eq. 1, threshold functions are c=0.436 for $\{x_n\}_{n=0}^{\infty}$ and $c' = 0.634$ for $\{y_n\}_{n=0}^{\infty}$ [19]. Thus, pseudorandom bit sequence $\{z_n\}_{n=0}^{\infty}$ can be defined as Eq. 6, that c and $c'$ are 0.436 and 0.634, respectively.

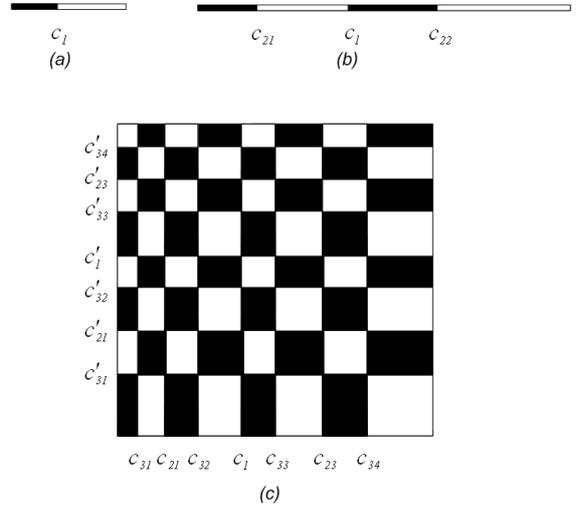

Fig 2: (a) threshold function of a block (k=1), (b) the generalized threshold function of $2^1$,(k=2) blocks in one-dimension, (c) the generalized threshold function of $2^2$,(k=3) blocks in two-dimension, the boxes of 0 and 1 be showed black and white colors, respectively.

### 3.3 Analysis of randomness of number sequences

We have survey the randomness and uniformity of the several bit sequences of large size, generated by the proposed PRNGs for different sets of control parameter and initial *conditions of the chaotic maps. Here, we show the results for $2^{20}$ sized bit sequences corresponding to the following parameter values of the four sets:*

$$\begin{cases} A = (0.2, 0.6, 4, 3.99997, 3) \\ B = (0.7, 0.3, 3.99998, 3.99996, 4) \\ C = (0.4, 0.8, 0.436, 0.634, 3) \\ D = (0.6, 0.3, 0.436, 0.634, 4) \end{cases}$$

*For convenience, these four sets are designated as:*

$$\begin{cases} A, B = (x, y, r_x, r_y, k) \\ C, D = (x, y, \alpha_x, \alpha_y, k) \end{cases}$$

*that A and B are related control parameter values of the PRNG based on segmentation and C and D are related control parameter values of the PRNG based on self-similarity. We have used MATLAB 7.10.0 (R2010a) running program in a personal computer with a Core i3 3.1GHz intel, 4GB memory and 500GB hard-disk capacity. The average time used for generating random bit sequences with size of $2^{20}$ bits is shorter than 0.3 s.*

*We discuss in the following paragraph of this Section the result and conclusions of our study of the different statistical tests to observe the randomness and uniformity of the bit sequences generated by the proposed PRNG. The US NIST statistical test suite provides 15 statistical tests to detect deviations of a bit sequence from randomness. A statistical test is formulated to test a null hypothesis which states that the sequence being tested is random. There is also an alternative hypothesis which states that the sequence is not random. For each test, there is an associated reference distribution (typically normal distribution or $\chi^2$ distribution), based on which a P-value is computed from the bit sequence. If the P-value is greater than a predefined threshold $\alpha$ which is also called significance level, then the sequence would be considered to be random with a confidence of $1-\alpha$, and the sequence passes the test successfully. Otherwise, the sequence fails this test. A P-value of zero indicates that the sequence appears to be completely non-random, and the larger the P-value is, the closer a sequence to a perfect random sequence. In our experiment, we set $\alpha$ to its default value 0.01, which means a sequence passed the test is considered as random with 99% confidence. Before presenting the test results of our proposed three approaches, we would first introduce all 15 statistical tests briefly as follows. A more detailed description for those tests could be found in [2].*

*The frequency test (FT), the runs test (RT) and the cumulative sum test (CST) are recommended that each sequence to be tested consist of a minimum of $10^2$ bits (i.e., $n \geq 10^2$). The frequency Test within a Block (FTB) is recommended that each sequence to be tested consist of a minimum of $M \times N$ bits (i.e., $n \geq MN$). The block size M should be selected such that $M \geq 20$ and $N < 10^2$. The discrete fourier transform test (DFTT) is recommended that each sequence to be tested consist of a minimum of $10^3$ bits (i.e., $n \geq 10^3$). The approximate entropy test (AET) is recommended that each sequence to be tested consist of a minimum of $2^{12}$ bits (i.e., $n \geq 2^{12}$). The test for the longest run of ones in a block (LROBT) is recommended that each sequence to be tested consist of a minimum of 6272 bits for M=128. The binary matrix rank test (BMRT) is recommended that each sequence to be tested consist of a minimum of $10^5$ bits (i.e., $n \geq 10^5$). The non-overlapping template matching test (NTMT), the overlapping template matching test (OTMT), the maurer's universal statistical test (MUST), the linear complexity test (LCT), the serial test (ST), the random excursions test (RET) and the random excursions variant test (REVT) are recommended that each sequence to be tested consist of a minimum of $2^{20}$ bits (i.e., $n \geq 2^{20}$).*

*The NIST suite tests were performed on four bit sequences, each containing $2^{20}$ bits. The P-value as well as final results obtained from the NIST suite for four different sets are given in Table 1. The proposed PRNGs successfully pass all randomness tests of NIST suite. According to [1], we can conclude that the data generated by these PRNGs are random.*

*Table 1. Shows the P-values obtained from NIST suite for fifteen different tests. The P-values are obtained for four different sets of parameters for each test.*

| NIST Tests | A | B | C | D |
|---|---|---|---|---|
| FT | 0.10128 | 0.95950 | 0.87737 | 0.50042 |
| FBT | 0.84189 | 0.64194 | 0.38602 | 0.92200 |
| RT | 0.97716 | 0.96884 | 0.85741 | 0.13400 |
| LROBT | 0.80721 | 0.03414 | 0.74711 | 0.48042 |
| RBMRT | 0.56967 | 0.77062 | 0.54269 | 0.52658 |
| DFTT | 0.07511 | 0.44408 | 0.78391 | 0.05936 |
| ATMT | PASS | PASS | PASS | PASS |
| PTMT | 0.21344 | 0.78374 | 0.94976 | 0.68709 |
| MUST | 0.13269 | 0.21346 | 0.40399 | 0.17531 |
| LCT | 0.29186 | 0.69918 | 0.83787 | 0.93722 |
| ST (P1) | 0.07947 | 0.41921 | 0.17617 | 0.05515 |
| (P2) | 0.48049 | 0.44903 | 0.40804 | 0.01233 |
| AET | 0.19412 | 0.33766 | 0.70768 | 0.58916 |
| CST (FORWARD) | 0.19653 | 0.79288 | 0.97455 | 0.68534 |
| (REVERSE) | 0.06508 | 0.83809 | 0.88538 | 0.57830 |
| RET | PASS | PASS | PASS | PASS |
| REVT | PASS | PASS | PASS | PASS |

## 4. THE PROPOSED ENCRYPTION SCHEME AND SECURITY ANALYSIS

*In this section, we introduce the encryption scheme based on the proposed PRNGs and then their security is discussed.*

### 4.1 Encryption scheme based on pseudorandom bit padding

*In the proposed scheme, we create a method to encrypt the image using bits padding. To consider a gray scale image with the size of $M \times N$. The steps of the encryption are shown below:*

- *Step 1: Generate $8 \times M \times N$ pseudo-random number sequence using one of the proposed PRNGs.*

- *Step 2: Transform the image into $8 \times M \times N$ bit sequence (image sequence).*
- *Step 3: Perform the XOR operation between the image sequence and the pseudo-random bit sequence to form the cipher sequence.*
- *Step 4: Transform the cipher sequence into image matrix I (ciphered image).*
- *Step 5: Divide the matrix I into four parts, uniformly. Move the odd columns with the even columns between the two parts in the main diagonal and between the other two parts, respectively.*
- *Step 6: Divide the matrix I into four parts, uniformly. Move the odd rows with the even rows between the two parts in the main diagonal and between the other two parts, respectively.*

*An indexed image of an 'Albert Einstein' sized $256 \times 256$ (see Fig. 3(a)) is used as a plain image and the encrypted images are shown in Figs. 3. The grey scale histograms are given in Figs. 4. The Figs. 4(b), 4(c), 4(d) and 4(e) show uniformity in distribution of grey scale of the encrypted images. In addition, the average pixel intensity for plain image is 98.92, and for encrypted images are 127.60, 127.69, 127.32, 127.55, respectively.*

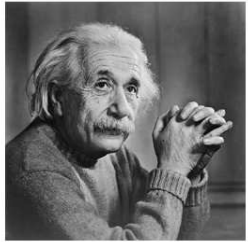
(a) Plain image

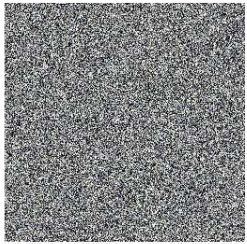
(b) Encrypted image from example of A

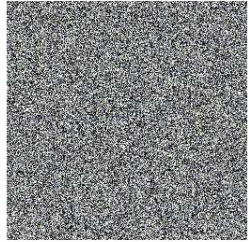
(c) Encrypted image from example of B

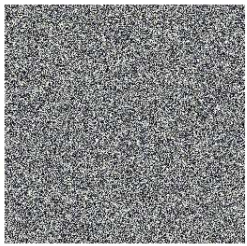
(d) Encrypted image from example of C

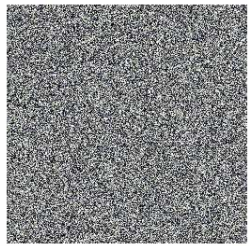
(e) Encrypted image from example of D

*Fig 3: Images of test results.*

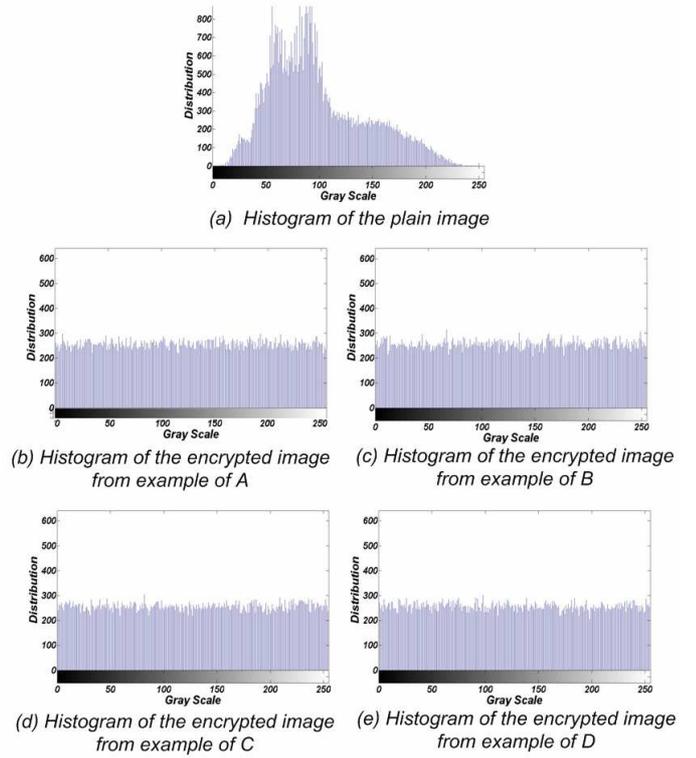

*Fig 4: Histograms of images.*

## 4.2 Analysis of security of the proposed encryption scheme

*Security is a major intransitive of a cryptosystem. Here, a complete analysis is made on the security of the cryptosystem. We have tried to explain that this cipher image is sufficiently secure against various cryptographical attacks, as shown below:*

### 4.2.1 Key space analysis

*The Key space size is the total number of different keys that can be used in the encryption [21]. Security issue is the size of the key space. If it is not large enough, the attackers may guess the image with brute-force attack. If the precision is $10^{-14}$, the size of key spaces for initial conditions and control parameters of the self-similarity method and segmentation method are $2^{186}$ and $2^{206}$, respectively. These sizes are large enough to defeat brute-force by any super computer today.*

### 4.2.2 Correlation Coefficient analysis

*The statistical analysis has been performed on the encrypted images from examples of A, B, C, D. This is shown by a test of the correlation between two adjacent pixels in plain image and encrypted image. We randomly select 2000 pairs of two-adjacent pixels (in vertical, horizontal, and diagonal direction) from plain images and ciphered images, and calculate the correlation coefficients, respectively by using the following two equations (see Table 2 and Fig. 5) [20,21]:*

$$Cov(x,y) = \frac{1}{N}\sum_{i=1}^{N}(x_i - E(x))(y_i - E(y)),$$

$$r_{xy} = \frac{Cov(x,y)}{(D(x))^{\frac{1}{2}}(D(y))^{\frac{1}{2}}}$$

Where

$$E(x) = \frac{1}{N}\sum_{i=1}^{N}(x_i), \quad D(y) = \frac{1}{N}\sum_{i=1}^{N}(x_i - E(x))^2.$$

Where, $E(x)$ is the estimation of mathematical expectations of $x$, $D(x)$ is the estimation of variance of $x$, and $Cov(x,y)$ is the estimation of covariance between $x$ and $y$, where $x$ and $y$ are grey scale values of two adjacent pixels in the image.

Table 2: Correlation coefficients of two adjacent pixels in the plain image and the ciphered images of examples of A, B, C, D

| Direction | Plain Image | Ciphered Image | | | |
|---|---|---|---|---|---|
| | | A | B | C | D |
| Horizontal | 0.9341 | 0.0026 | 0.0019 | 0.0015 | 0.0004 |
| Vertical | 0.9634 | 0.0087 | 0.0095 | 0.0021 | 0.0046 |
| Diagonal | 0.9402 | 0.0036 | 0.0038 | 0.0029 | 0.0024 |

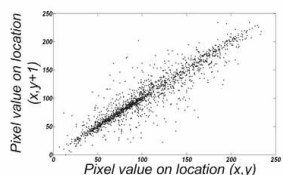
(a) The plain image

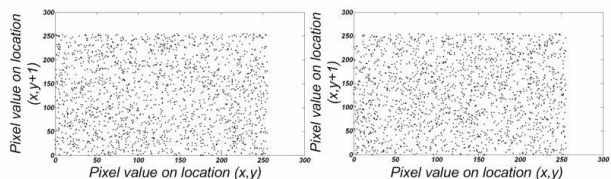
(b) The ciphered image of example of A    (c) The ciphered image of example of B

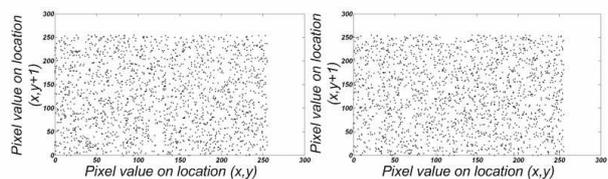
(d) The ciphered image of example of C    (e) The ciphered image of example of D

Fig 5: Correlation distributions of two horizontally adjacent pixels in the plain image and the ciphered images.

### 4.2.3 Differential attack

Attackers try to find out a relationship between the plain image and the cipher image, by studying how differences in an input can affect the resultant difference at the output in an attempt to derive the key [22]. Trying to make a slight change such as modifying one pixel of the plain image, attacker observes the change of the cipher image [22]. To test the influence of one pixel change on the whole encrypted image by the proposed scheme, two common measures are used:

Number of Pixels Change Rate (NPCR) stands for the number of pixels change rate while, one pixel of plain image is changed. Unified Average Changing Intensity (UACI) measures the average intensity of differences between the plain image and ciphered image. The NPCR and The UACI, are used to test the influence of one pixel change on the whole image encrypted by the proposed scheme and can be defined as following:

$$NPCR = \frac{\sum_{i,j} D(i,j)}{W \times H} \times 100\%$$

$$UACI = \frac{1}{W \times H}\left[\sum_{i,j}\frac{C_1(i,j) - C_2(i,j)}{255}\right] \times 100\%$$

where $W$ and $H$ are the width and height of $C_1$ or $C_2$. $C_1$ and $C_2$ are two ciphered images, whose corresponding original images have only one pixel difference and also have the same size. The $C_1(i,j)$ and $C_2(i,j)$ are grey-scale values of the pixels at grid $(i,j)$. The $D(i,j)$ determined by $C_1(i,j)$ and $C_2(i,j)$. If $C_1(i,j) = C_2(i,j)$, then, $D(i,j) = 1$; otherwise, $D(i,j) = 0$. We have done some tests on the proposed scheme (256 grey scale image of size $256 \times 256$) to find out the extent of change produced by one pixel change in the plain image (see Table 3). The results demonstrate that the proposed scheme can survive differential attack.

Table 3: Results of the differential attack in the ciphered images of examples of A, B, C, D.

| | A | B | C | D |
|---|---|---|---|---|
| NPCR | 42% | 48% | 45% | 38% |
| UACI | 38% | 33% | 31% | 39% |

## 5. CONCLUSION

We have proposed a chaotic encryption scheme based on pseudorandom bit padding that the bits be generated by the novel generalized threshold function (segmentation and self-similarity) methods. The security of the cipher images of this scheme is evaluated by the key space analysis, the correlation of two adjacent pixels and differential attack. The distributions of the ciphered images are very close to the uniform distribution, which can well protect the information of the image to withstand the statistical attack. We suggest the use of these methods to design other secure cryptosystems.